\documentstyle[aps,amsfonts,preprint,floats,epsfig]{revtex}
\begin{document}
\unitlength 1mm
\title{Asymptotic behaviour for critical slowing-down random walks}
\author{Yves Elskens\cite{bylineYE}}
\address{Equipe turbulence plasma de l'UMR 6633
         CNRS--Université de Provence,
              \\ case 321, Centre de Saint-J\'er{\^o}me,
              F-13397 Marseille Cedex 13         \\
         \vskip0.5cm
         {\em NONLINEAR SCIENCE : DYNAMICS \& STOCHASTICITY}
         \\
    {\em invited talk on the occasion of the 60th birthday of Gr{\'e}goire Nicolis}
         \\
         Universit{\'e} Libre de Bruxelles, June 30 - July 3, 1999
         }
\date{preprint TP99.21 to appear in {\bf Journal of Statistical Physics 101} (2000) 397-404} 
\maketitle

\narrowtext

\begin{abstract}
  The jump processes $W(t)$ on $[0,\infty[$ with transitions
$w \to \alpha w$ at rate $bw^\beta$ ($0 \leq \alpha \leq 1$,
$b>0$, $\beta>0$) are considered. Their moments are shown to decay
not faster than algebraically for $t \to \infty$,
and an equilibrium probability density
is found for a rescaled process $U = (t + \kappa)^{-\beta} W$.
A corresponding birth process is discussed.

PACS numbers: \\%
\noindent 02.50.Ey Stochastic processes               \\%
\noindent 05.20.Dd Kinetic theory                     \\%
\noindent 83.70.Fn Granular solids                    \\%
\noindent 05.40.Fb Random walks and Levy flights      \\%

AMS 1991 MSC :  \\%
\noindent 60C18 Self-similar stochastic processes     \\%
\noindent 82C23 Exactly solvable dynamic models
                (time-dependent statistical physics)  \\%
\noindent 70F35 Collisions
\end{abstract}

\section{Introduction}

Random walks with absorbing states occur quite commonly. Here we
focus on the asymptotic behaviour of random walks suffering
critical slowing down on approaching such a state
\cite{BS1,ElskensCom,ErnstCom,BS2}.

Consider a markovian random walk $W$ on $[0,+\infty[$ with
continuous time $t$. The value of $W$ is multiplied at random
times by a given factor $0 \leq \alpha < 1$, and successive jumps
occur independently, with a waiting time between them distributed
according to an exponential law with parameter $b W^\beta$, where
$b>0$, $\beta > 0$. By introducing a new time $t' = b t$, a new
variable $W' = W^\beta$ and a new constant $\alpha' =
\alpha^\beta$, one reduces this model to the case $\beta=1$, $b=1$
with no loss of generality. A characteristic parameter related to
$\alpha$ is the $e$-folding number of jumps
\begin{equation}
  \nu = - 1 / \ln(\alpha)
\label{nu}
\end{equation}
such that, after $c \nu$ jumps, $W$ has been reduced by a factor
$\alpha^{c \nu} = e^{-c}$.

Denoting by $f(w,t)$ the probability distribution function of $W$,
the evolution equation for $f$ reads
\begin{eqnarray}
  \partial_t f(w,t) = - w f(w,t) + \alpha^{-2}w f(\alpha^{-1}w,t)
  & {\hskip1cm}
  & \text{ for } \alpha > 0
  \label{dft}
  \\
  \partial_t f(w,t) = - w f(w,t) + \delta (w) x_1(t)
  & {\hskip1cm}
  & \text{ for } \alpha = 0
\label{dft0}
\end{eqnarray}
If they exist, the moments $x_k(t) = \langle W^k \rangle =
\int_0^\infty w^k f(w,t) dw$ satisfy the hierarchy
\begin{equation}
  \dot x_k = - (1 - \alpha^k) x_{k+1}
\label{xk}
\end{equation}
with the dot denoting derivation with respect to time $t$.
We assume that ${\mathbb P}(W(0)>0)=1$.

As this random walk can move only towards the origin, we are
interested in its asymptotic approach to $0$. Equations
(\ref{dft})-(\ref{xk}) indicate that $W$ and $t^{-1}$ are
dimensionally homogeneous, and no absolute time scale is available
in these equations.

Physically speaking, this random walk appears naturally in the
kinetic theory of inelastic systems. Up to dimensional constants,
if $W$ is the modulus of the velocity of a particle suffering
inelastic collisions (with restitution factor $\alpha$) with fixed
colliders, our process describes the particle's velocity slowing
down. For an assembly of such particles, the resulting process is
thus inelastic cooling. However, further physical analysis of
actual granular systems leads to a more complex process that our
simple random walk \cite{ElskensCom,ErnstCom}.

\section{Evolution for $W$}

The linear hierarchy (\ref{xk}) has a straightforward formal
solution \cite{BS1,BS2}
\begin{equation}
  x_k(t) = \sum_{l=0}^\infty {{(-t)^l} \over {l!}} \
           {{[k+l-1]!} \over {[k-1]!}} \
           x_{k+l}(0)
\label{xkt}
\end{equation}
with the q-factorial notation $[k]! = \prod_{n=1}^k \Bigl((1 -
\alpha^n) / (1 - \alpha)\Bigr)$. Note that $\lim_{\alpha \to 1}
[k]! = k!$ and that $0 < \lim_{k \to \infty} (1-\alpha)^k [k]! <
1$ for $0 < \alpha < 1$.

If $x_k(0) < M^k$ for some $M>0$ uniformly with respect to $k \geq
0$, then (\ref{xkt}) implies that all moments $x_k(t)$ are entire
functions of $t$, bounded by $M^k e^{(1-\alpha) M |t|}$. Also, if
$0 \leq f(w,0) < c \exp(-c'w^\gamma) \ \forall w \in ]0, \infty[$
for some $c>0$, $c'>0$, $\gamma > 1$, moments $x_k$ ($k \in
{\mathbb N}$) are entire functions of time. Conversely, hierarchy
(\ref{xk}) is ill-defined if some initial moments are infinite,
though (\ref{dft}) and (\ref{dft0}) are well-defined in this case
too. These bounds are useful to discuss analytic properties of
(\ref{xkt}), but they are poor for positive times, as $0 < x_k(t)
< x_k(0)$ obviously for $t>0$.

Consider now the evolution equation (\ref{dft}). By linearity and
scale invariance, its general solution reads
\begin{equation}
  f(w,t) = \int_0^\infty f(v,0) Q({w \over v}, v t) \frac{dv}{v}
\label{fQ}
\end{equation}
where $Q$ is the solution of (\ref{dft}) for Dirac initial
data $Q(w,0) = \delta(w-1)$.

For $\alpha = 0$, %
$Q(w,t) = e^{-t} \delta(w-1) + (1- e^{-t}) \delta(w)$, so that for
$w>0$, $f(w,t)=e^{-wt}f(w,0)$. The moments follow readily, $x_k(t)
= k! t^{-k-1} f(0,0) + {o}(t^{-k-1})$, and the surviving fraction
of the initial population ${\mathbb P}(W(t)>0) = t^{-1} f(0,0) +
{o}(t^{-1})$, if $f(w,0)$ is smooth near $w=0$.

For $\alpha > 0$, the representation $Q(w,t) =
\sum_{n=0}^\infty q_n(t) \delta(w - \alpha^n)$ yields coefficients
$q_n$ by induction. Indeed, $\dot q_0 = - q_0$, $\dot q_n = -
\alpha^n q_n + \alpha^{n-1} q_{n-1}$, with initial data $q_0(0) =
1$, $q_n(0) = 0$ for $n>0$. The Laplace transforms ${\hat q}_n(s) =
\int_0^\infty q_n(t) e^{-st} dt$ follow for $\Re (s) >0$~:
\begin{eqnarray}
  {\hat q}_0(s) &=& (1 + s)^{-1}
  \\
  {\hat q}_n(s) = {{\alpha^{n-1}} \over {s + \alpha^n}} \  {\hat q}_{n-1}
        &=& {1 \over {s+1}}
            \prod_{k=1}^n { {\alpha^{k-1}} \over {s + \alpha^k}}
\end{eqnarray}
The accumulation of poles $s_n = - \alpha^n$ to the origin
reflects the critical slowing down of the process.

The Laplace transforms ${\hat X}_k(s)$ of moments
$X_k(t) = \sum_n q_n(t) \alpha^{nk}$ have singular expansions for
$s \to 0$.
In particular, $(1+s){\hat X}_1 (s) = 1+{\hat X}_1 (s/\alpha)$
for $s>0$, which admits the solution
${\hat X}_1 (s) = A(s) \ln s + B(s)$
with entire functions $A(s) = \sum_{n=0}^\infty a_n s^n$,
$B(s) = \sum_{n=0}^\infty b_n s^n$ near $s=0$~:
$a_n = a_0 \prod_{k=1}^n (\alpha^{-k}-1)^{-1}$,
$b_n = (b_{n-1} \alpha^n + a_n \ln \alpha)/(1 - \alpha^n)$,
$a_0 = \nu$.

\section{Rescaled process}

As the natural time scale for the evolution of $W$ is
$1/W$, consider the rescaled variables
\begin{eqnarray}
   \tau &=& \ln (1 + t/\kappa)
   \label{deftau}
   \\
   U(\tau) &=& (t + \kappa) W(t)
   \label{defU}
\end{eqnarray}
where the characteristic time $\kappa$ is adapted to the specific
initial data $f$. The moments $y_k$ of $U$ and its probability
density $h$ are
\begin{eqnarray}
  y_k (t) &=& (t + \kappa)^k x_k(t)
  \label{defy}
  \\
  h(u,\tau) &=& (t + \kappa) f(w,t)
  \label{defh}
\end{eqnarray}
and for $0 < \alpha < 1$ obey the equations
\begin{eqnarray}
  y_k' &=& k y_k - (1 - \alpha^k) y_{k+1}
  \label{eqy}
  \\
  y_k(0) &=& x_k(0) \kappa^k
  \label{iny}
  \\
  \partial_\tau h &=& L h(u,\tau) + \alpha^{-2} u h(\alpha^{-1} u, \tau)
  \label{eqh}
  \\
  h(u,0) &=& \kappa^{-1} f(u/\kappa, 0)
  \label{inh}
\end{eqnarray}
where $Lh(u,\tau) = - u \partial_u h(u,\tau) - (u+1) h(u,\tau)$
and the prime denotes derivation with respect to $\tau$. It is
easily seen that $U$ is a stationary Markov process, ergodic on
the half line $]0,\infty[$. It jumps down by a factor $\alpha$
with rate $u$ and drifts upwards along exponential characteristics
($u'=u$).

As $U$ is ergodic, one finds a stationary density $h^{\rm eq}(u)$
and its moments $y_k^{\rm eq}$ :
\begin{equation}
  y_{k+1}^{eq} =  y_1^{\rm eq} \prod_{l=1}^k \frac{l}{(1-\alpha^l)}
               =  y_1^{\rm eq} {\frac {k!} {[k]!}} (1 - \alpha)^{-k}
\label{ykeq}
\end{equation}
Moreover, as (\ref{eqy})-(\ref{iny}) hold for all
$k \in {\mathbb R}$ and $y_0=1$ by normalisation,
one finds $y_1^{\rm eq}= \nu = -1/\ln(\alpha)$ in the limit $k\to 0$.
Induction for negative $k$ shows that the stationary solution
of (\ref{eqy}) has also finite moments for all $-\infty < k < +\infty$
(diverging for $|k| \to \infty$).

\begin{figure}
\centerline{
  \psfig{figure=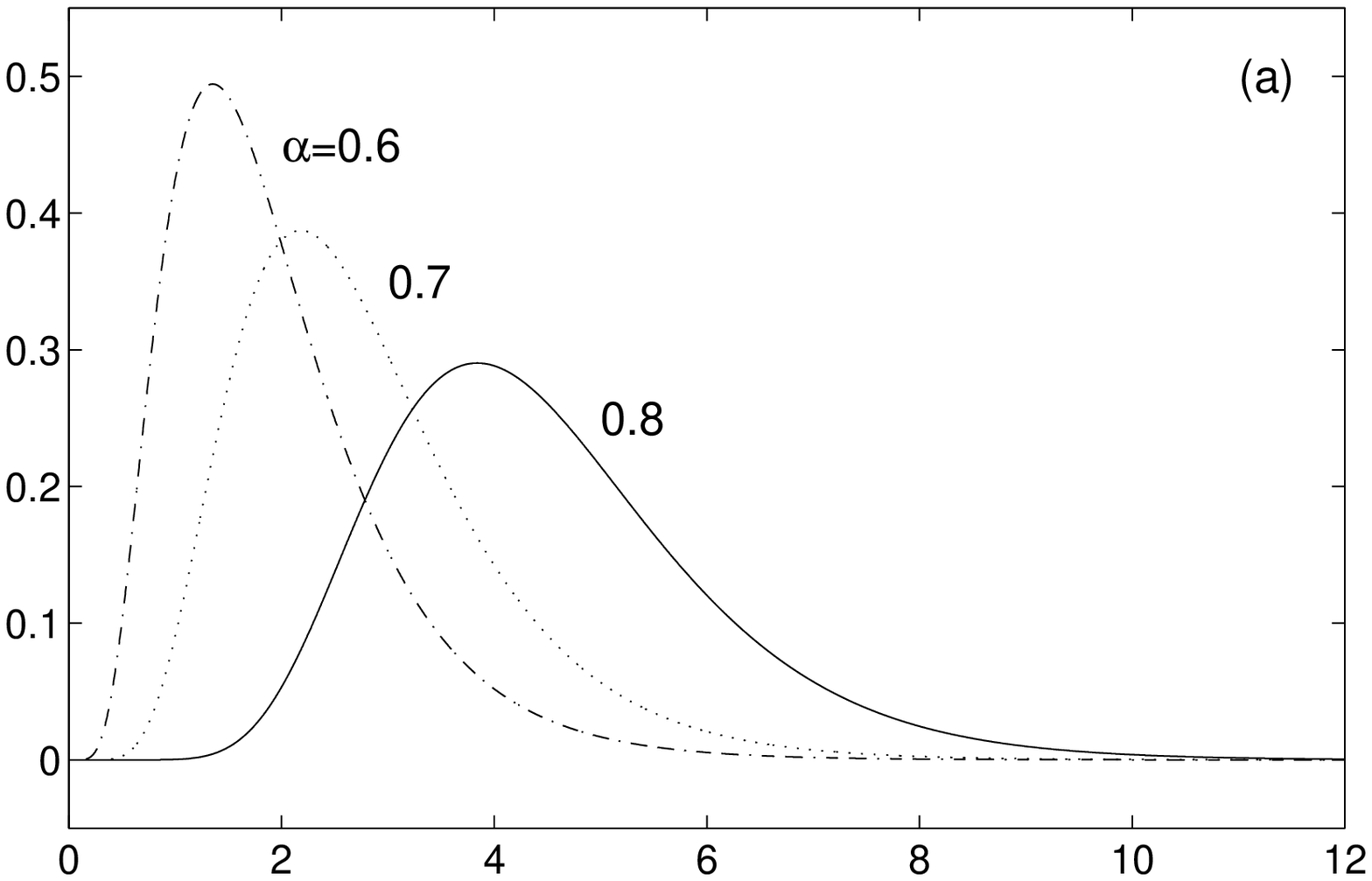,width=7cm,height=4.5cm}
  \hskip1cm %
  \psfig{figure=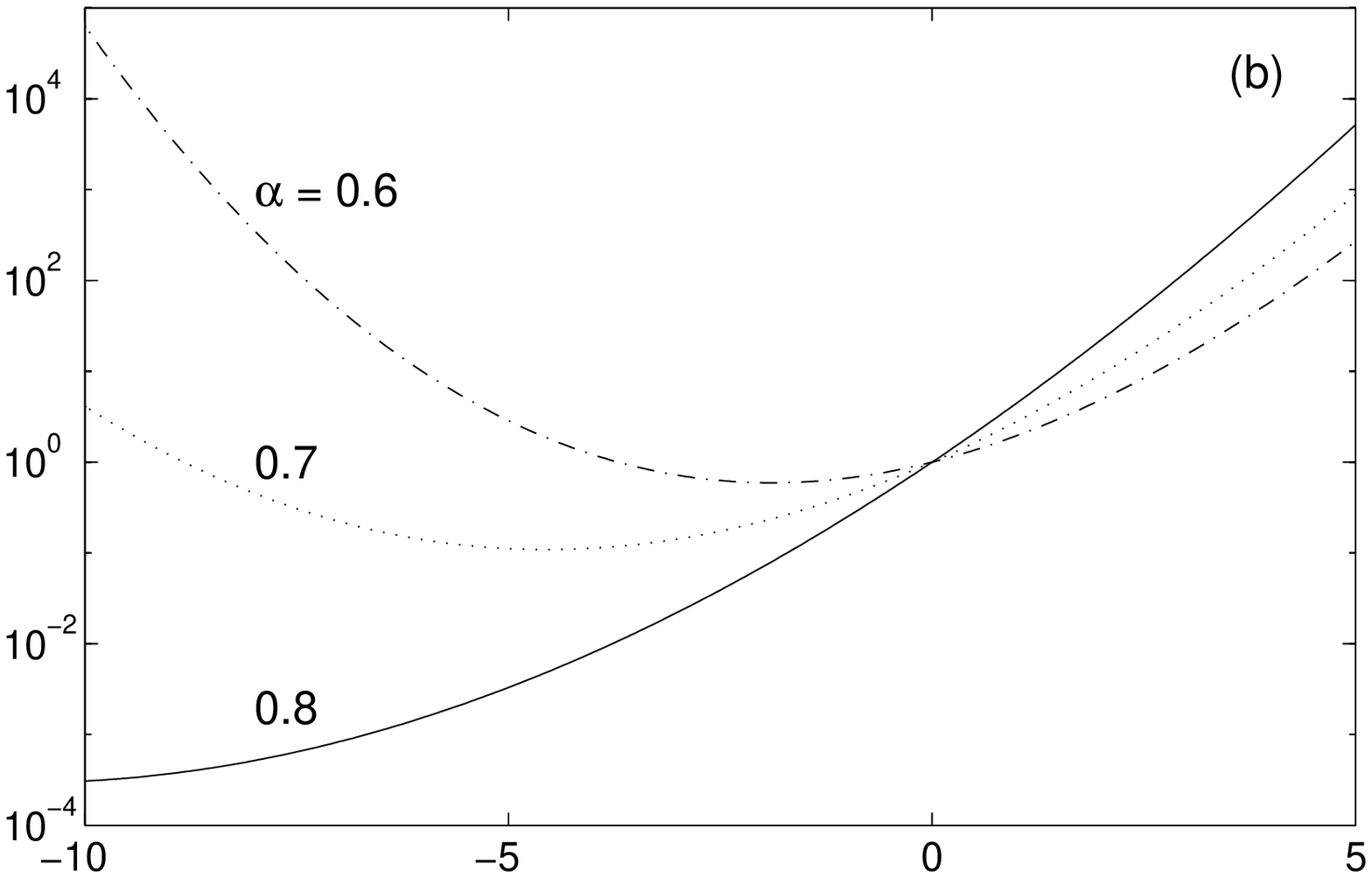,width=7cm,height=4.5cm}
  }
  \vskip1cm
\caption{ For $\alpha = 0.8$ (solid line), $0.7$ (dots) and $0.6$
(dash-dots)~: (a) invariant density $h^{\rm eq}$ vs $u$~; (b)
moments $y_k^{\rm eq}$ vs $k$ (in semi-log scale).}
\label{figheq}
\end{figure}

A series for $h^{\rm eq}$ is found in the form
\begin{equation}
  h^{\rm eq} (u) = \sum_{m=0}^\infty \eta_m u^{-1} e^{-u/\alpha^m}
\label{heq}
\end{equation}
with coefficients $\eta_m = - \eta_{m-1} \alpha^{-1} (\alpha^{-m} -1)^{-1}$
(note that $\eta_m \sim \alpha^{m(m-1)/2}$ for $m \to \infty$).
This series converges in the half plane  $\Re (u) >0$,
with an essential singularity at $u=0$. At that point,
all derivatives of $h^{\rm eq}$ vanish, in agreement with the finiteness of
its moments. Figure \ref{figheq} displays this stationary density and its moments.


\begin{figure}
\centerline{
  \psfig{figure=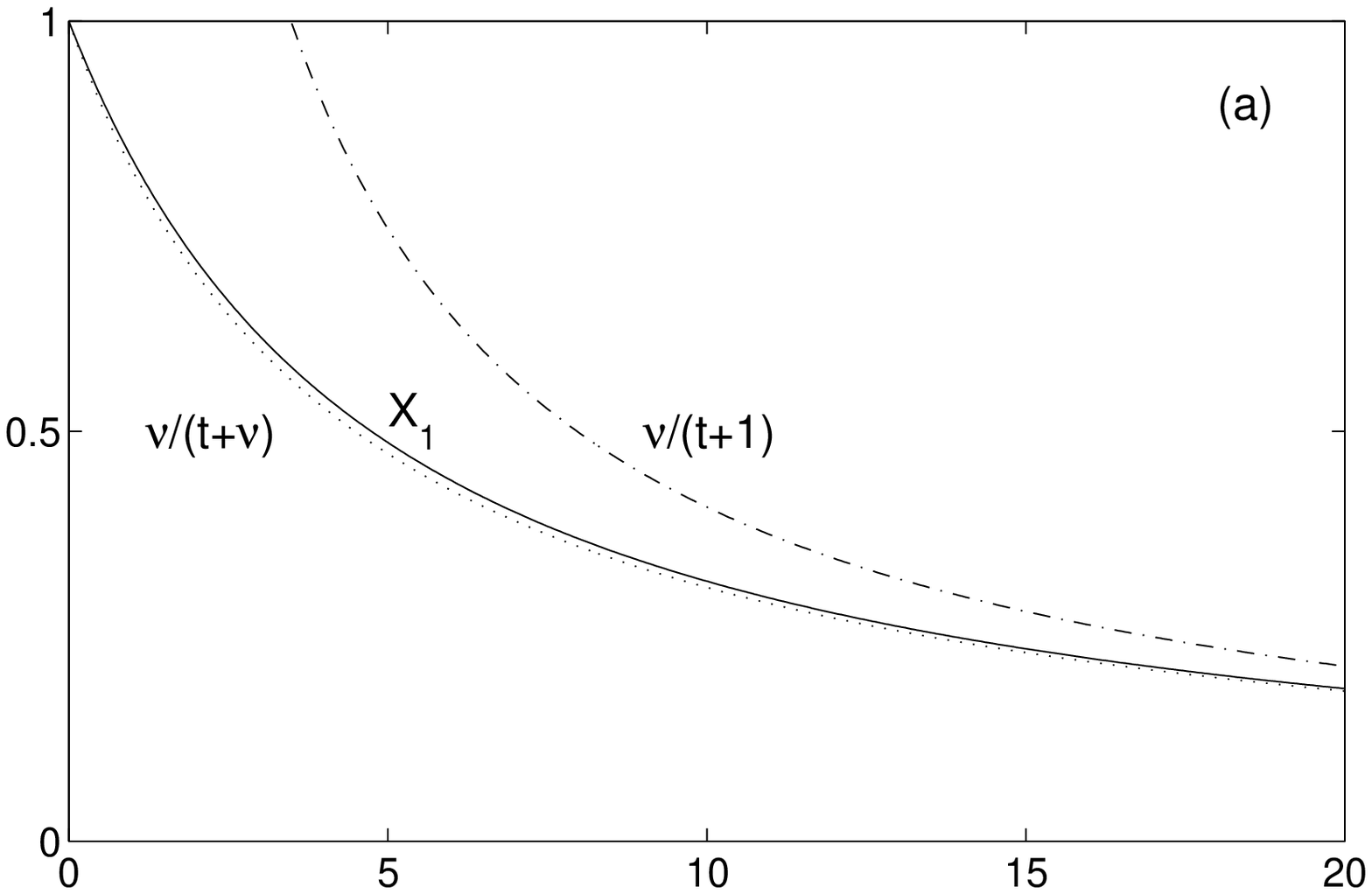,width=7cm,height=4.5cm}
  \hskip1cm %
  \psfig{figure=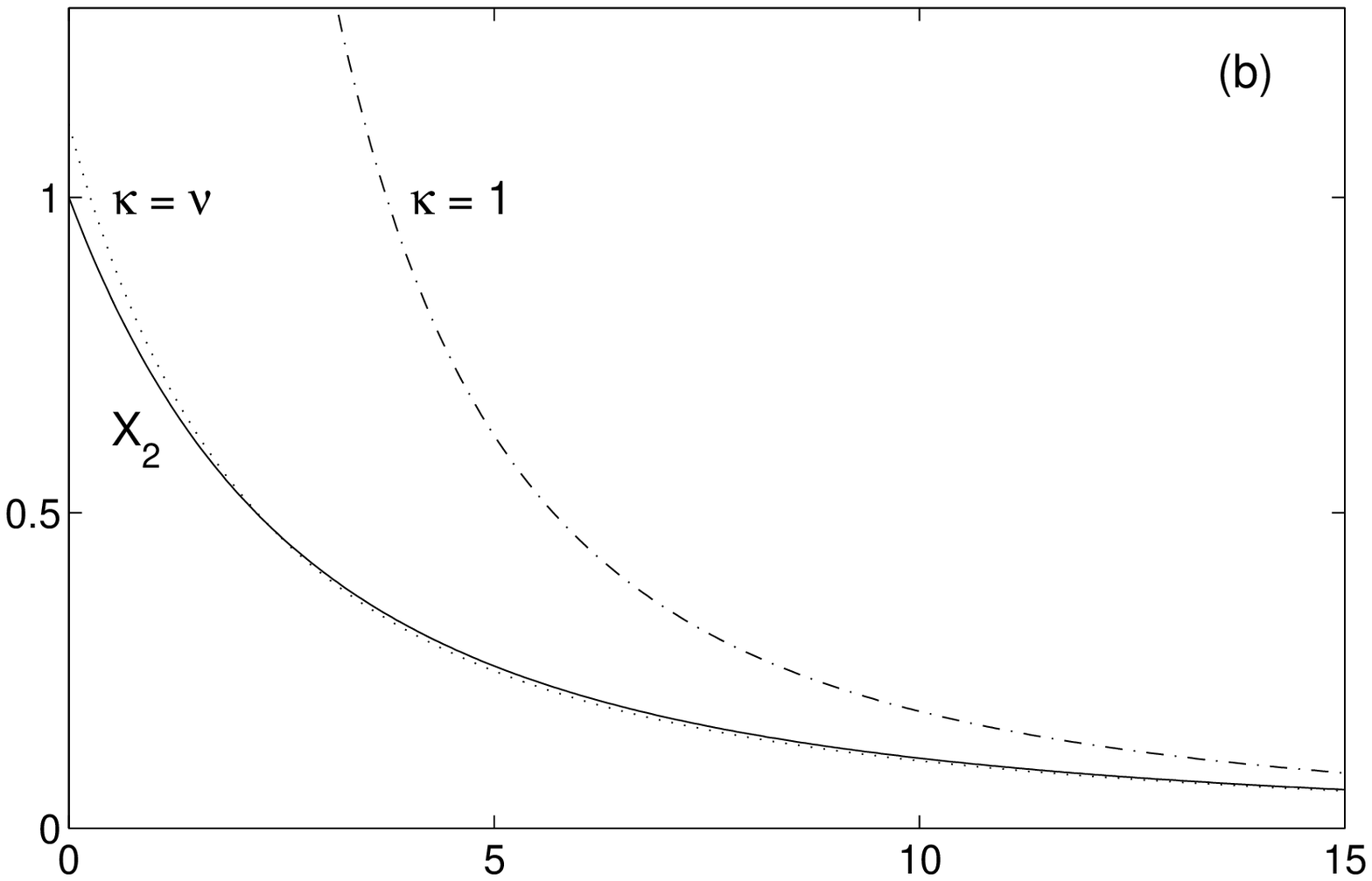,width=7cm,height=4.5cm}
  }
  \vskip1cm
\caption{ Expectations $X_k (t)$ for the kernel $Q$ with $\alpha =
0.8$ and (a) $k=1$, (b) $k=2$. Solid line~: direct sum of series
(\ref{xkt}) with $x_k(0)=1 \ \forall k$~; thin lines~:
asymptotically leading approximation $(t+\kappa)^{-k} y_k^{\rm
eq}$ with $\kappa = 1$ (dots) and $\kappa = \nu$ (dash-dots).}
\label{figx1}
\end{figure}

While initial distributions $h(u,0)$ relax to the stationary
density $h^{\rm eq}$, the distribution of the original variable
$W$ approaches \cite{indata} this profile, up to the rescaling by
$(t + \kappa)$. In particular, Figure \ref{figx1} displays the
first two moments of the Green function $Q$ and the leading term
$(t + \kappa)^{-k} y_k^{\rm eq}$ in its asymptotic expansion for
$t \to \infty$. Two choices for time scale $\kappa$ are compared~:
$\kappa=1$ may seem natural in view of initial data ${\mathbb
P}(W(0)=1)=1$, but the choice $\kappa = y_1^{\rm eq} /X_1(0) =
\nu$ has the advantage that the first moment satisfies $y_1(0) =
y_1^{\rm eq}$ and, indeed, the relaxation of $(t +
\kappa)Q(u/(t+\kappa),t)$ to $h^{\rm eq}$ is almost unnoticed on
these lower moments with the choice $\kappa = \nu$.


\section{Population dynamics interpretation}

One may also interpret our process as a birth (only) process for
the variable $Z = \ln(\kappa W) / \ln \alpha$ (with the same
characteristic time $\kappa$ as introduced in (\ref{deftau}) and
(\ref{defU})) : $Z$ increases by unit jumps, with a jump rate
$\kappa^{-1}\alpha^Z$. This description suggests that $Z$ would
describe e.g. a self-inhibiting single-species population growth
process, where all present individuals of the species are
cooperating to produce one more individual.

It is easily seen that, for the population corresponding to the
asymptotic distribution,
\begin{equation}
  \langle Z(t) \rangle
  \approx
  \nu \ln(1+t/\kappa)
  - \nu \langle \ln U \rangle^{\rm eq}
\label{asZ}
\end{equation}
where $\langle \ln U \rangle^{\rm eq} = \lim_{k \to 0} d y_k^{\rm
eq} /dk$. The population grows to infinity, logarithmically in
time, as does the solution $z = \nu \ln (a + t/\kappa) - \nu \ln
\nu$ (with integration constant $a$) to the rate equation $\dot z
= \kappa^{-1} \alpha^z$ corresponding to the birth process.

\section{Concluding remarks}

The long-time behaviour of moments $x_k$ in the case
$f(w,0)$ does not vanish on a neighbourhood of $w=0$
raises interesting questions, as the density $h^{\rm eq}$ is flat
at the origin. Relaxation of $(t+\kappa)f(u/(t+\kappa),t)$ is likely to be algebraic,
and the competition between the scale factor $(t+\kappa)^{-1}$ and the
relaxation of $h$ may lead to non-universality in the asymptotics.

One may also consider that the absorption at the origin is a coarse
description of the relevant physical processes. In this respect, one may
balance the drift towards zero by various simple processes :
\begin{enumerate}
  \item{adding a diffusion term to (\ref{dft}), with reflecting boundary
      condition ($\partial_w f(0,t) = 0$) \cite{Lambiotte}~;}
  \item{imposing a constant acceleration (with $W$ interpreted as a velocity)
      between jumps, as in the one-dimensional inelastic Lorentz gas model
      \cite{Martin}.}
\end{enumerate}
Analytic investigations of these models reveal rich behaviours and varied
structures of equilibrium distributions \cite{Lambiotte,Martin}.

\acknowledgements

It is a pleasure to thank L. Brenig, R. Lambiotte, J. Piasecki,
J.M. Salazar Cruz and participants to CEMRACS'99 (CIRM, Luminy)
for discussions, and E. Andjel for suggesting to find a rescaled
process. The author is pleased to recall his first exposure to
random walks in physics and chemical kinetics by G. Nicolis, and
that this work initiated from his invitation by G. Nicolis to
universit{\'e} libre de Bruxelles last year.

%


\begin{references}
\bibitem{indata} {This is clear for initial measures $f$ with compact
    support in $]0,\infty[$ (bounded away from the origin),
    such as the Dirac distribution. Asymptotics
    with $f(0,0)>0$ or with $f(w,0)$ decaying too slowly for $w \to \infty$
    may be slower.}
\end{references}
\end{document}